\documentclass[11pt,epsf]{article}

\usepackage{epsfig}
\usepackage{amssymb}
\usepackage{graphicx}
\usepackage{color}
\usepackage{subfigure}
\usepackage{mathtools}

\makeatletter

\@addtoreset{equation}{section} \makeatother
\def\a{\alpha}

\def\g{\gamma}

\def\d{\delta}

\def\ep{\epsilon}

\def\ph{\phi}
\def\Ph{\Phi}

\def\l{\lambda}
\def\L{\Lambda}
\def\m{\mu}
\def\n{\nu}
\def\th{\theta}

\def\r{\rho}
\def\s{\sigma}

\def\ta{\tau}
\def\o{\omega}
\def\O{\Omega}

\def\w{\wedge}

\def\p{\partial}

\def\mcp{\mathcal{P}}
\def\mcq{\mathcal{Q}}
\def\mct{\mathcal{T}}

\def\lt{\left}
\def\rt{\right}

\def\nn{\nonumber}

\DeclarePairedDelimiter\abs{\lvert}{\rvert}%

\begin{document}

\begin{titlepage}
\title{\vskip -60pt
\vskip 20pt Generalized complex marginal deformation of pp-waves and
giant gravitons }
\author{
Sunyoung Shin\footnote{e-mail : shin@theor.jinr.ru}}
\date{}
\maketitle \vspace{-1.0cm}
\begin{center}
~~~
\it Bogoliubov Laboratory of Theoretical Physics, JINR, 141980
Dubna, Moscow region, Russia
~~~\\
~~~\\
\end{center}

\thispagestyle{empty}

\begin{abstract}
We present the Penrose limits of a complex marginal deformation of
$AdS_5\times S^5$, which incorporates the $SL(2,\mathbb{R})$
symmetry of type IIB theory, along the $(J,0,0)$ geodesic and along
the $(J,J,J)$ geodesic. We discuss giant gravitons on the deformed
$(J,0,0)$ pp-wave background.
\end{abstract}

\end{titlepage}
\newpage


\section{Introduction}
\setcounter{equation}{0}
The marginal deformation \cite{Leigh:1995ep} introduces phases in
the superpotential which breaks the $SO(6)_R$ R-symmetry group to
its $U(1)\times U(1) \times U(1)_R$ Cartan subgroup. In the gravity
side \cite{Lunin:2005jy}, the $U(1) \times U(1)$ non-R-symmetry maps
to a two-torus. The dual geometry is obtained by applying an
$SL(2,\mathbb{R})$ transformation which acts on the K\"{a}hler
modulus of the corresponding two-torus or equivalently a TsT
(T-duality, shift, T-duality) transformation. The phases in the
gauge theory can be complexified. In the dual geometry, it
corresponds to a specific $SL(3,\mathbb{R})$ transformation which
consists of the $SL(2,\mathbb{R})$ transformation and an S-duality
transformation $SL(2,\mathbb{R})_s$ or equivalently an STsTS
(S-duality, T-duality, shift, T-duality, S-duality) transformation
\cite{Lunin:2005jy,Frolov:2005dj}.\footnote{$\g$ is used for the
$SL(2,\mathbb{R})$ transformation and $\s$ is used for the
$SL(2,\mathbb{R})_s$ transformation. Both are real parameters with
unit period.} The three-parameter generalization is proposed as a
dual geometry to a non-supersymmetric marginal deformation of
$\mathcal{N}=4$ super Yang-Mills theory \cite{Frolov:2005dj}.

The charges of chiral superfields under the $U(1)\times U(1)$
symmetry in the gauge theory corresponds to the angular momenta
along the two-torus in the dual geometry. In terms of the angle
coordinates $(\ph_1,\ph_2,\ph_3)$ of $S^5$, there are four possible
BPS geodesics, $(J_{\ph_1},J_{\ph_2},J_{\ph_3})\sim (J,0,0)$,
$(0,J,0)$, $(0,0,J)$ and $(J,J,J)$. The Penrose limit along the
first three geodesics and the Penrose limit along the fourth
geodesic are two distinct pp-waves. The pp-waves are discussed in
\cite{Niarchos:2002fc,Lunin:2005jy,deMelloKoch:2005vq,Avramis:2007wb}.

A point graviton which has an angular momentum about the sphere of
$AdS_m \times S^n$ blows up into a spherical brane
\cite{McGreevy:2000cw}. A giant graviton is a spherical
$(n-2)$-brane which wraps a part of $S^n$. A dual giant graviton is
a spherical $(m-2)$-brane which wraps a spatial part of $AdS_m$.
Both are BPS objects, which have the same quantum numbers as the
Kaluza-Klein mode of the point graviton
\cite{Grisaru:2000zn,Hashimoto:2000zp}. Giant gravitons in the
Penrose limit of $AdS_5\times S^5$ are studied in
\cite{Takayanagi:2002nv}.

Giant gravitons on the three-parameter non-supersymmetric background
\cite{Frolov:2005dj} are discussed in
\cite{deMelloKoch:2005jg,Pirrone:2006iq}. It is shown in
\cite{Pirrone:2006iq} that the (dual) giant gravitons do not depend
on the deformation parameters $\g_i$, $(i=1,2,3)$. (Dual) giant
gravitons in the supersymmetric deformation are obtained by setting
$\g_i=\g$. D3-brane (dual) giant gravitons and D5-brane dual giant
gravitons on $\g$-deformed $AdS_5\times S^5$ are discussed in
\cite{Imeroni:2006rb}. Giant gravitons in the Penrose limits of
marginally deformed $AdS_5\times S^5$ along the $(J,0,0)$ geodesic
and along the $(J,J,J)$ geodesic are considered in
\cite{Hamilton:2006ri,Avramis:2007wb}. It is shown in
\cite{Hamilton:2006ri} that the giant graviton on the deformed
$(J,0,0)$ pp-wave is independent of the deformation parameter $\g$
and energetically degenerate with the Kaluza-Klein point graviton
whereas the giant graviton on the deformed $(J,J,J)$ pp-wave does
not retain its round three-sphere shape. In \cite{Avramis:2007wb},
the Penrose limits of the complex marginal deformation of
$AdS_5\times S^5$ along the $(J,0,0)$ geodesic and along the
$(J,J,J)$ geodesic are studied. Giant gravitons and dual giant
gravitons are discussed on the deformed $(J,0,0)$ pp-wave. It is
shown that the giant gravitons are not energetically degenerate with
the point graviton and exist only up to a critical value of $\s$.
They are energetically unfavorable but nevertheless perturbatively
stable.

In this work, we study the Penrose limits of complex marginal
deformation of $AdS_5\times S^5$, which incorporates the
$SL(2,\mathbb{R})$ symmetry of type IIB theory and observe giant
gravitons on the deformed $(J,0,0)$ pp-wave background. In section
\ref{sec:ppwave}, we review the generalized complex marginal
deformation of $AdS_5\times S^5$
\cite{Shin:2013oya,Dagvadorj:2014aja}, and present the pp-wave
geometries which are obtained by taking the Penrose limits along the
$(J,0,0)$ geodesic and along the $(J,J,J)$ geodesic. In section
\ref{sec:J00}, we study the giant graviton solution on the pp-wave
background and check the stability by observing small fluctuations
about the solution. In section \ref{sec:diss}, we summarize our
results.
%
\section{Generalized complex marginal deformation}\label{sec:ppwave}
\setcounter{equation}{0}

The Lunin-Maldacena $SL(3,\mathbb{R})$ transformation, which
generates the gravity dual of the complex marginal deformation
\cite{Leigh:1995ep,Lunin:2005jy} is
\begin{eqnarray}
\L_{LM}^T=\lt(
\begin{array}{ccc}
1   &   0   &   0 \\
\g  &   1   &   \s  \\
0   &   0   &   1
\end{array}
\rt).
\end{eqnarray}
The transformation can be generalized by an $SL(3,\mathbb{R})$
transformation
\begin{eqnarray} \label{eq:sl_matrix}
L=\lt(
\begin{array}{ccc}
L_{11} & 0 & L_{13}  \\
0      & 1 & 0       \\
L_{31} & 0 & L_{33}
\end{array}
\rt),~~\det{L}=1,
\end{eqnarray}
which corresponds to the $SL(2,\mathbb{R})$ symmetry of type IIB
supergravity. The $SL(3,\mathbb{R})$ transformation $L\L_{LM}^T$,
therefore produces a generalized complex marginal deformation
\cite{Lunin:2005jy,Shin:2013oya}.
We consider $AdS_5\times S^5$ defined by
\begin{eqnarray}
ds^2=R^2\lt[-dt^2\cosh^2\r+d\r^2+\sinh^2\r d\O_3^2+ \sum^3_{i=1}
d\m_i^2 + \sum^3_{i=1}\m_i^2d\ph_i^2 \rt], \nn
\end{eqnarray}
\begin{eqnarray} \label{eq:ads5s5}
&&\chi_0=\ta_1,~~e^{-\Phi_0}=\ta_2,~~B_2=0,~~C_2=0,\nn\\
&&C_4=4R^4e^{-\Phi_0}(\o_4+\o_1\w d\ph_1 \w d\ph_2 \w d\ph_3), \nn\\
&&F_5=4R^4e ^{\Phi_0}(\o_{AdS_5}+\o_{S^5}), \nn\\
&&\o_{AdS_5}=d\o_4,~~\o_{S^5}=d\o_1 \w d\ph_1 \w d\ph_2 \w d\ph_3,
\nn\\
&&d\o_1=\cos\a \sin^3\a \cos\th \sin\th d\a \w d\th, \nn\\
&&\m_1=\cos\a,~~\m_2=\sin\a \cos\th,~~\m_3=\sin\a \sin\th,
\end{eqnarray}
where $R$ is the radius of $AdS_5$ and the radius of $S^5$.
The complex marginal deformation of $AdS_5\times S^5$
\cite{Dagvadorj:2014aja} is
\begin{eqnarray} \label{eq:beta_ads1}
&&ds^2=R^2H^{1/2}\Big[-dt^2\cosh^2\r+d\r^2+\sinh^2\r
d\O_3^2+\sum_{i=1}^3\lt(d\m_i^2+G\m_i^2d\ph_i^2\rt)\nn\\
&&\hspace{1cm}
+G\mcp\m_1^2\m_2^2\m_3^2\lt(\sum_{i=1}^3d\ph_i\rt)^2\Big],
\nn\\
&&e^{\Phi}=\sqrt{G}H\ta_2^{-1},\nn\\
&&\chi=H^{-1}\lt(h+\ta_2^2\hat{\g}\hat{\s}g_0\rt),\nn\\
&&B_2=R^2G\mcq\o_2-4R^2\ta_2\hat{\s}\o_1\w\sum_{i=1}^3d\ph_i,
\nn\\
&&C_2=R^2G\mct\o_2-4R^2\ta_2\hat{\g}\o_1\w\sum_{i=1}^3d\ph_i,\nn\\
&&C_4=4R^4\ta_2\o_4+4R^4\ta_2G\Big[1-\hat{\s}\mct g_0\Big]\o_1\w
d\ph_1\w d\ph_2\w d\ph_3,\nn\\
&&F_5=4R^4\ta_2(\o_{AdS_5}+G\o_{S^5}),
\end{eqnarray}
where
\begin{eqnarray}\label{eq:pqt}
\mcp&=&\hat{\g}^2f-2\hat{\g}\hat{\s}h+\hat{\s}^2g, \nn\\
\mcq&=&\hat{\g}f-\hat{\s}h,   \nn\\
\mct&=&\hat{\g}h-\hat{\s}g,
\end{eqnarray}
\begin{eqnarray}
G^{-1}&=&1+\mcp g_0,
\nn\\
H&=&f+\ta_2^2\hat{\s}^2g_0, \nn\\
g_0&=&\m_1^2\m_2^2+\m_2^2\m_3^2+\m_3^2\m_1^2,\nn\\
\o_2&=&\m_1^2\m_2^2d\ph_1\w d\ph_2+\m_2^2\m_3^2d\ph_2\w
d\ph_3+\m_3^2\m_1^2d\ph_3\w d\ph_1,
\end{eqnarray}
and
\begin{eqnarray}\label{eq:fgh}
f&=&\lt(L_{33}+L_{13}\ta_1\rt)^2+{L_{13}}^2 \ta_2^2, \nn\\
g&=&\lt(L_{31}+L_{11}\ta_1\rt)^2+{L_{11}}^2 \ta_2^2, \nn\\
h&=&\lt(L_{33}+L_{13}\ta_1\rt)\lt(L_{31}+L_{11}\ta_1\rt)+L_{11}L_{13}
\ta_2^2.
\end{eqnarray}
The $SL(2,\mathbb{R})$ transformation (\ref{eq:sl_matrix}) can be
identified with torus parameters from an eleven dimensional
viewpoint. The parametrization considered in
\cite{Shin:2013oya,Dagvadorj:2014aja} is
\begin{eqnarray}
L_{11}=1,~~L_{13}=\frac{r_3}{R_1}\cos\xi,~~L_{31}=0,~~L_{33}=1,
\end{eqnarray}
with a constraint
\begin{eqnarray}
r_3=\frac{R_3}{\sin\xi}.
\end{eqnarray}
$R_i$, $(i=1,3)$ are the torus radii before the torus deformation
and $r_3$ is the torus radius of the third direction after the
deformation. $\xi$ is the intersection angle between the direction
along the first direction and the direction along the third
direction. The geometry can be simplified by identifying the
axion-dilaton coupling with the torus modulus of the rectangular
torus before the torus deformation as
\begin{eqnarray}
\ta=\ta_1+i\ta_2=il,~~l:=\frac{R_1}{R_3}.
\end{eqnarray}
The deformed $AdS_5\times S^5$ is \cite{Shin:2013oya}
\begin{eqnarray}\label{eq:simp_def}
ds^2&=&R^2\widetilde{H}^{1/2}\Big[-dt^2\cosh^2\r+d\r^2+\sinh^2\r
d\O_3^2+\sum_{i=1}^3\lt(d\m_i^2+\widetilde{G}\m_i^2d\ph_i^2\rt)
+9\widetilde{G}\widetilde{\mcp}\m_1^2\m_2^2\m_3^2d\psi^2\Big],\nn\\
e^{\Ph}&=&\sqrt{\widetilde{G}}\widetilde{H}l^{-1},\nn\\
\chi&=&\widetilde{H}^{-1}\lt(l\cot\xi+\hat{\g}\hat{\s}l^2g_0\rt),\nn\\
B_2&=&R^2\widetilde{G}\widetilde{\mcq}\o_2-4R^2\hat{\s}l\o_1\w\sum_{i=1}^3d\ph_i,\nn\\
C_2&=&R^2\widetilde{G}\widetilde{\mct}\o_2-4R^2\hat{\g}l\o_1\w\sum_{i=1}^3d\ph_i,\nn\\
C_4&=&4R^4l\o_4+4R^4l\tilde{G}(1-\hat{\s}\mct g_0)\o_1\w d\ph_1\w
d\ph_2\w d\ph_3,\nn\\
F_5&=&4R^4l(\o_{AdS_5}+\widetilde{G}\o_{S^5}),
\end{eqnarray}
where
\begin{eqnarray}
\widetilde{G}^{-1}&=&1+{\widetilde{\mcp}}g_0,\nn\\
\widetilde{H}&=&\csc^2\xi+\hat{\s}^2l^2g_0, \nn\\
{\widetilde{\mcp}}&=&\hat{\g}^2\csc^2\xi-2\hat{\g}\hat{\s}l\cot\xi+\hat{\s}^2l^2,
\nn\\
{\widetilde{\mcq}}&=&\hat{\g}\csc^2\xi-\hat{\s}l\cot\xi,\nn\\
{\widetilde{\mct}}&=&\hat{\g}l\cot\xi-\hat{\s}l^2.
\end{eqnarray}

We study the Penrose limits of (\ref{eq:beta_ads1}) along the
$(J,0,0)$ geodesic and along the $(J,J,J)$ geodesic. The
parametrization to take the Penrose limit along the $(J,0,0)$
geodesic is
\begin{eqnarray}
&&\Xi_1:=f,~~\r=\frac{y}{\Xi_1^{1/4}R},~~\a=\frac{r}{\Xi_1^{1/4}R}, \nn\\
&&t=x^++\frac{x^-}{2\Xi_1^{1/2}R^2},~~\ph_1=x^+-\frac{x^-}{2\Xi_1^{1/2}R^2},
\nn\\
&&r^2=\sum_{i=1}^4(x^i)^2,~~y^2= \sum_{a=5}^8(x^a)^2.
\end{eqnarray}
By taking $R \rightarrow \infty$, we obtain the pp-wave geometry
\begin{eqnarray} \label{eq:geoj00_1}
&&ds^2=-2dx^+dx^--\big[y^2+(1+\mcp) r^2\big]{(dx^+)}^2
+dr^2+r^2 d\tilde{\O}_3^2+dy^2+y^2d\O_3^2,\nn\\
&&e^{\Ph}=\Xi_1 \ta_2^{-1},\nn\\
&&B_2=\frac{r^2}{\Xi_1^{1/2}}\mathcal{Q}(\cos^2\theta dx^+\w
d\ph_2-\sin^2\th dx^+ \w d\ph_3),\nn\\
&&C_2=\frac{r^2}{\Xi_1^{1/2}}\mathcal{T}(\cos^2\theta dx^+\w
d\ph_2-\sin^2\th dx^+ \w d\ph_3),\nn\\
&&C_4=-\frac{\ta_2}{\Xi_1}(y^4dx^+\w d\O_3+r^4dx^+\w d\tilde{\O}_3),
\end{eqnarray}
where
\begin{eqnarray}\label{eq:S3_ads}
&&d\tilde{\O}^2_3=d\th^2+\cos^2\th d\ph_2^2+\sin^2\th d\ph_3^2,
\nn\\
&&d\tilde{\O}_3=\cos\th\sin\th d\th \w d\ph_2 \w d\ph_3.
\end{eqnarray}

The parametrization to take the Penrose limit along the $(J,J,J)$
geodesic \cite{Dagvadorj:2014aja} is
\begin{eqnarray}
&& \Xi_2:=f+\frac{1}{3}\hat{\s}^2\ta_2^2, ~~\th_0=\frac{\pi}{4},~~\a_0=\arccos(\frac{1}{\sqrt{3}}), \nn\\
&&\a=\a_0-\frac{x^2}{\Xi_2^{1/4}R},
~~\th=\th_0+\sqrt{\frac{3}{2}}\frac{x^1}{\Xi_2^{1/4}R},~\r=\frac{y}{\Xi_2^{1/4}R},\nn\\
&&\varphi^1=\sqrt{\frac{3+\mcp}{2}}\frac{1}{\Xi_2^{1/4}R}\lt(x^3-\frac{1}{\sqrt{3}}x^4\rt),~~
\varphi^2=\sqrt{\frac{2(3+\mcp)}{3}}\frac{x^4}{\Xi_2^{1/4}R},\nn\\
&&t=x^++\frac{1}{2\Xi_2^{1/2}R^2}x^-,~~\psi=x^+-\frac{1}{2\Xi_2^{1/2}R^2}x^-,
\end{eqnarray}
where the spherical coordinates and the torus coordinates are
related by
\begin{eqnarray}
\ph_1=\psi-\varphi_2,~~\ph_2=\psi+\varphi_1+\varphi_2,~~\ph_3=\psi-\varphi_1.
\end{eqnarray}
By taking $R \rightarrow \infty$ and shifting the coordinate $x^-$
as $x^- \rightarrow x^-
-\frac{\sqrt{3}}{\sqrt{3+\mathcal{P}}}(x^1x^3+x^2x^4)$, we obtain
the pp-wave geometry in the homogeneous plane wave form
\cite{Papadopoulos:2002bg} \footnote{$\o_1$ in (\ref{eq:ads5s5}) is
solved as
$\o_1=\frac{1}{R^2}\Xi_2^{-1/2}\lt[\lt(\frac{\sqrt{3}}{9}-\zeta\rt)x^1dx^2-\zeta
x^2dx^1\rt]+\mathcal{O}(R^{-3})$. $\zeta=\frac{\sqrt{3}}{18}$ is
chosen, which is consistent with \cite{Hamilton:2006ri}.}
\begin{eqnarray}\label{eq:geojjj_1}
&&ds^2=-2dx^+dx^-+\frac{2\sqrt{3}}{\sqrt{3+\mcp}}(x^3dx^1+x^4dx^2-x^1dx^3-x^2dx^4)dx^++\sum_{I=1}^8(dx^I)^2
\nn\\
&&\hspace{1cm}-\lt[\sum_{a=5}^8(x^a)^2+\frac{4\mathcal{P}}{3+\mathcal{P}}\lt((x^1)^2+(x^2)^2\rt)\rt](dx^+)^2,
\nn\\
&& e^{\Ph}=\sqrt{\frac{3}{3+\mcp}}\Xi_2\ta_2^{-1},\nn\\
&& B_2=\frac{\mathcal{Q}}{\sqrt{3}}\Xi_2^{-1/2}dx^3\w dx^4 +
\frac{2\mathcal{Q}}{\sqrt{3+\mathcal{P}}}\Xi_2^{-1/2}dx^+\w(x^2dx^3-x^1dx^4)\nn\\
&&\hspace{1cm}-\frac{2\sqrt{3}}{3}\hat{\s}\ta_2\Xi_2^{-1/2}dx^+\w
(x^2dx^1-x^1dx^2), \nn\\
&& C_2=\frac{\mathcal{T}}{\sqrt{3}}\Xi_2^{-1/2}dx^3\w
dx^4+\frac{2\mathcal{T}}{\sqrt{3+\mathcal{P}}}\Xi_2^{-1/2}dx^+\w(x^2dx^3-x^1dx^4)\nn\\
&&\hspace{1cm}-\frac{2\sqrt{3}}{3}\hat{\g}\ta_2\Xi_2^{-1/2}dx^+\w (
x^2dx^1-x^1dx^2), \nn\\
&&C_4=4R^4 \ta_2 \o_4 \nn\\
&&\hspace{1cm}+2\ta_2\Xi^{-1}\lt(1-\frac{1}{3}\mathcal{T}
\hat{\s}\rt)dx^+ \w \lt(x^2 dx^1\w dx^3 \w dx^4 - x^1  dx^2 \w dx^3
\w dx^4\rt).
\end{eqnarray}
%
\section{Giant graviton on the deformed pp-wave} \label{sec:J00}
\setcounter{equation}{0}
%
We study giant gravitons on the deformed pp-wave
(\ref{eq:geoj00_1}). A static gauge for a brane which wraps the
$(\th,\ph_2,\ph_3)$ directions is
\begin{eqnarray}
\s^0=\ta, ~~ \s^1=\th,~~  \s^2=\ph_2,~~  \s^3=\ph_3,
\end{eqnarray}
and
\begin{eqnarray}
X^+=\l \ta,~~ X^-=\m\ta.
\end{eqnarray}
The fields on the three-sphere (\ref{eq:S3_ads}) can be
parameterized as
\begin{eqnarray}
&&X^1=r\cos\theta\cos\ph_2,~~
X^2=r\sin\theta\cos\ph_3,  \nn\\
&&X^3=r\cos\theta\sin\ph_2,~~ X^4=r\sin\theta\sin\ph_3.
\end{eqnarray}
We turn off the fields on $AdS_5$
\begin{eqnarray}
X^a=0,~~(a=5,6,7,8).
\end{eqnarray}

A D3-brane is described by the Dirac-Born-Infeld action and the
Wess-Zumino term\footnote{We choose the minus sign for the
Wess-Zumino term as it is done in \cite{Imeroni:2006rb} since it is
consistent with the conventions of
\cite{Lunin:2005jy,Frolov:2005dj}.}
\begin{eqnarray}
S&=&S_{\mathrm{DBI}}+S_{WZ} \nn\\
&=&-T_3\int d^4\s e^{-\Ph}\sqrt{-\det \mathrm{P}[g-B_2]}-T_3\int
\sum \mathrm{P} [C_q \w e^{-B_2}].
\end{eqnarray}
$\mathrm{P}$ denotes the pullback of the spacetime field to the
brane worldvolume. The D3-brane action in the deformed geometry
(\ref{eq:geoj00_1}) is
\begin{eqnarray}
S&=&-T_3\int d^4\s
\frac{\ta_2}{\Xi_1}\sqrt{-\det\mathrm{P}[g-B_2]}-T_3\int
\mathrm{P}[C_4] \nn\\
&=&-\frac{2\pi^2T_3\ta_2}{\Xi_1}\int d\ta
\lt[r^3\sqrt{2\l\m+\l^2r^2(1+\hat{\s}^2\ta_2^2\Xi_1^{-1})}-\l
r^4\rt].
\end{eqnarray}
The action does not depend on $\g$ while it depends on $\s$ as well
as $\ta_1$ and $\ta_2$.

The lightcone momentum\footnote{The conjugate momenta are defined by
$P_\pm=\frac{\d L}{\d (\p_\ta X^\pm)}$. The upper index and the
lower index are related by $P^\pm=-P_\mp$.} of the D3-brane is
\begin{eqnarray} \label{eq:lm_j00}
P^+=-\frac{\d L}{\d \m}=\frac{M\l
r^3}{\Xi_1\sqrt{2\l\m+\l^2r^2(1+\hat{\s}^2\ta_2^2\Xi_1^{-1})}}\,,
\end{eqnarray}
and the lightcone Hamiltonian is
\begin{eqnarray}
P^-=H_{lc}=-\frac{\d L}{\d \l}=\frac{Mr^3}{\Xi_1}\Big[\frac{\m+\l
r^2(1+\hat{\s}^2\ta_2^2\Xi_1^{-1})}{\sqrt{2\l\m+\l^2r^2(1+\hat{\s}^2\ta_2^2\Xi_1^{-1})}}-r\Big],
\end{eqnarray}
where $M:=2\pi^2\ta_2T_3$.
The lightcone Hamiltonian can be written as
\begin{eqnarray}
H_{lc} \sim \frac{M^2r^6}{2\Xi_1^2
P^+}+\frac{P^+(1+\hat{\s}^2\ta_2^2\Xi_1^{-1})r^2}{2}
-\frac{Mr^4}{\Xi_1}.
\end{eqnarray}
For $0 \leq \hat{\s} < \frac{\sqrt{\Xi_1}}{\sqrt{3}\ta_2}$, the
Hamiltonian is extremized at
\begin{eqnarray}\label{eq:radii_j00}
r_0=0,~r_\pm=\sqrt{\frac{P^+\Xi_1}{3M}\lt(2\pm\sqrt{1-3\hat{\s}^2\ta_2^2\Xi_1^{-1}}\rt)}.
\end{eqnarray}
The lightcone Hamiltonian has local minima at $r=r_0$ and $r=r_+$,
and a local maximum at $r=r_-$. The radii do not depend on $\g$
while they depend on $\s$ as well as the axion-dilaton parameters
$\ta_1$ and $\ta_2$. The corresponding lightcone energies are
\begin{eqnarray}
E_0&=&0, \nn\\
E_\pm&=&\frac{(P^+)^2\Xi_1}{27M}\lt[1+9\hat{\s}^2\ta_2^2\Xi_1^{-1}\mp\lt(1-3\hat{\s}^2\ta_2^2\Xi_1^{-1}\rt)^{3/2}\rt].
\end{eqnarray}
For $\hat{\s}=0$, we have $E_+=E_0$, i.e., the giant graviton is
degenerate with the point graviton. For
$0<\hat{\s}<\frac{\sqrt{\Xi_1}}{\sqrt{3}\ta_2}$, we have $E_+>E_0$,
i.e., the degeneracy is lifted. The energy of the point graviton is
less than the energy of the giant graviton. Therefore the giant
graviton becomes energetically unfavorable. For
$\hat{\s}=\frac{\sqrt{\Xi_1}}{\sqrt{3}\ta_2}$, we have $r_+=r_-$ and
$E_+=E_-$, i.e., there is a saddle point at $r=r_\pm$. The lightcone
Hamiltonian has one minimum at $r=r_0$. For $\hat{\s}
> \frac{\sqrt{\Xi_1}}{\sqrt{3}\ta_2}$, the giant graviton
disappears. The result is consistent with the results of
\cite{Hamilton:2006ri,Avramis:2007wb}.

\begin{figure}[h!]
\begin{center}
\subfigure{
\includegraphics[angle=0,width=0.5\textwidth]{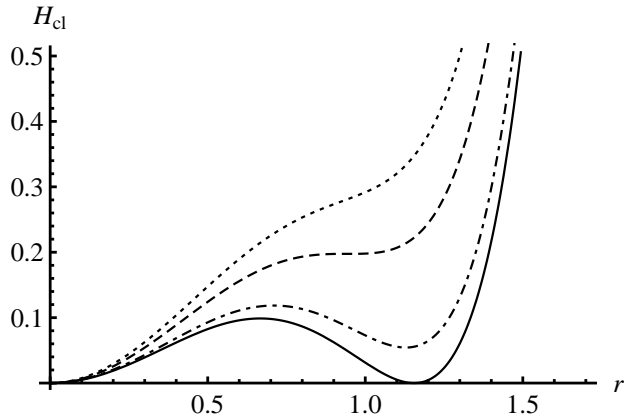}}
\vspace{0cm} \caption{Lightcone Hamiltonian with $\xi=\frac{\pi}{3}$
as a function of $r$. $\hat{\s}l=0$ (solid), $\hat{\s}l=\frac{1}{3}$
(dot-dashed), $\hat{\s}l=\frac{2}{3}$ (dashed),
$\hat{\s}l=\frac{2.5}{3}$ (dotted), $\Xi_1=\csc^2\xi$, $\ta_2=l$,
$M=1$ and $P^+=1$.}\label{fig:j00a}
\end{center}
\end{figure}

\begin{figure}[h!]
\begin{center}
\subfigure{
\includegraphics[angle=0,width=0.5\textwidth]{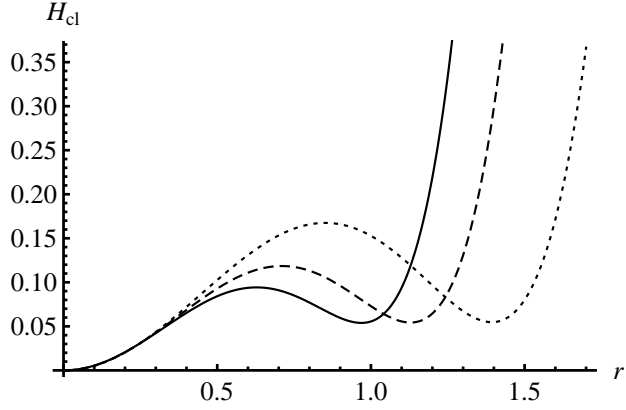}}
\vspace{0cm} \caption{Lightcone Hamiltonian with
$\hat{\s}l=\frac{1}{3}$ as a function of $r$.
$\xi=\frac{\pi}{2}$(solid), $\xi=\frac{\pi}{3}$ (dashed),
$\xi=\frac{\pi}{4}$ (dotted), $\Xi_1=\csc^2\xi$, $\ta_2=l$, $M=1$
and $P^+=1$.} \label{fig:j00b}
\end{center}
\end{figure}

The lightcone Hamiltonian in the Penrose limit of
(\ref{eq:simp_def}) is obtained by substituting $\Xi_1=\csc^2\xi$
and $\ta=\ta_1+i\ta_2=il$. The lightcone Hamiltonian for
$\xi=\frac{\pi}{3}$ in the units of $M=1$ and $P^+=1$ is plotted in
Figure \ref{fig:j00a}. It is qualitatively the same as the one
plotted in \cite{Avramis:2007wb}. The lightcone Hamiltonian for
$\hat{\s} l=\frac{1}{3}$ in the units of $M=1$ and $P^+=1$ is
plotted in Figure \ref{fig:j00b}. As the intersection angle $\xi$
decreases, the minimum value of the lightcone Hamiltonian increases.
%
%
%
%

We examine the spectrum of small fluctuations  about the giant
graviton solution following the method of
\cite{Hamilton:2006ri,Avramis:2007wb,Das:2000st}. We fix the
lightcone coordinates as
\begin{eqnarray}
X^+=\ta,~~X^-=\n\ta+\ep \d x^-.
\end{eqnarray}
The ansatz for the perturbed configuration is
\begin{eqnarray}
&&r=r_0+\ep \d r, \nn\\
&&X^1=r\cos\th\cos\ph_2,~~X^2=r\sin\th\cos\ph_3, \nn\\
&&X^3=r\cos\th\sin\ph_2,~~X^4=r\sin\th\sin\ph_3, \nn\\
&&X^a=\ep \d x^a,~~ (a=5,\cdots,8).
\end{eqnarray}
The components of the pullback $D_{\m\n}=\mathrm{P}[g-B_2]_{\m\n}$
up to the second order of $\ep$ are
\begin{eqnarray}
D_{\ta\ta}&=&-2\n-(1+\mcp)r_0^2+\ep\big[-2\p_\ta\d x^- -2(1+\mcp)r_0
\d r\big]\nn\\&&+\ep^2\big[-\sum_a(\d x^a)^2-(1+\mcp)\d
r^2+\sum_{I=i+a}(\p_\ta\d x^I)^2\big], \nn
\end{eqnarray}
\begin{eqnarray}
D_{\ta\th}=D_{\th\ta}=-\ep\p_{\th}\d x^-+\ep^2\sum_{I=i+a}(\p_\ta \d
x^I)(\p_\th \d x^I), \nn
\end{eqnarray}
\begin{eqnarray}
D_{\ta\ph_2 /\ph_2\ta}&=&\mp \Xi_1^{-1/2} \mcq r_0^2 \cos^2\th + \ep
[-\p_{\ph_2} \d x^- \mp 2 \Xi_1^{-1/2} \mcq r_0 \cos^2\th \d r] \nn\\
&&+\ep^2[\sum_{I=i+a}\p_\ta \d x^I\p_{\ph_2} \d x^I \mp \Xi_1^{-1/2}
\mcq  \cos^2\th \d r^2],
\end{eqnarray}
\begin{eqnarray}
D_{\ta\ph_3 /\ph_3\ta}&=&\pm \Xi_1^{-1/2} \mcq  r_0^2 \sin^2\th +\ep
[-\p_{\ph_3}\d x^- \pm 2  \Xi_1^{-1/2} \mcq r_0   \sin^2\th \d r]\nn\\
&&+\ep^2[\sum_{I=i+a}\p_\ta \d x^I \p_{\ph_3} \d x^I \pm
\Xi_1^{-1/2} \mcq \sin^2\th \d r^2], \nn
\end{eqnarray}
\begin{eqnarray}
D_{ij}=r_0^2g_{ij}+2\ep r_0 g_{ij}\d r+\ep^2\lt[g_{ij}\d
r^2+\sum_{I=1}^8(\p_i \d x^I)(\p_j \d x^I)\rt],
~~(i,j=\th,\ph_2,\ph_3),\nn
\end{eqnarray}
where the metric $g_{ij}$ is defined as
\begin{eqnarray}
g_{ij}=\lt(
\begin{array}{ccc}
1  &  0  &   0 \\
0  & \cos^2\th  &   0   \\
0  &  0  &  \sin^2\th
\end{array}
\rt).
\end{eqnarray}

The D3-brane action is
\begin{eqnarray} \label{eq:act_pert}
\mathcal{S}&=&\mathcal{S}_{\mathrm{DBI}}+\mathcal{S}_{\mathrm{WZ}} \nn\\
&=& -T_3 \int d^4\s e^{-\Ph}\sqrt{-\det \mathrm{P}[G-B]} - T_3 \int
\mathrm{P}[C_4-C_2\w B_2] \nn\\
&=& -T_3\Xi_1^{-1} \ta_2  \int d\ta d^3\s \sqrt{\abs{g}}
r_0^3\lt\{\sqrt{2\n+r_0^2 u}-r_0\rt\} \nn\\
&&-\ep T_3 \Xi_1^{-1}\ta_2  \int d\ta d^3\s \sqrt{\abs{g}}
\frac{r_0^3}{\sqrt{2\n+r_0^2u}} \lt\{\p_\ta \d x^- +\frac{2\d
r}{r_0} \lt[3\n+2r_0^2 u-2r_0\sqrt{2\n+r_0^2 u}\rt] \rt\} \nn\\
&&-\ep^2T_3\Xi_1^{-1}\ta_2 \int d\ta d^3\s
\frac{r_0}{2\sqrt{2\n+r_0^2u}} \Big\{\sqrt{\abs{g}}\Big[30\n+28
r_0^2 u
-\frac{(6\n+4r_0^2u)^2}{2\n+r_0^2u} \nn\\
&&-12 r_0 \sqrt{2\n+r_0^2u} \Big] \d r^2 +\sqrt{\abs{g}}
r_0^2\sum_a(\d x^a)^2  -\sum_{I} \d x^I \Big[(2\n+r_0^2
u)\p_i(\sqrt{|g|}g^{ij}\p_j)
\nn\\
&&+(\p_{\ph_2}-\p_{\ph_3})(\sqrt{\abs{g}}\Xi_1^{-1} \mcq^2 r_0^2
)(\p_{\ph_2}-\p_{\ph_3})-r_0^2\p_\ta\lt(\sqrt{\abs{g}}\p_\ta\rt)\Big]\d
x^I \nn\\
&&+\sqrt{\abs{g}}\frac{4(3\n+r_0^2 u)}{2\n+r_0^2u} r_0 \d r \d_\ta
\d
x^- - \d x^- \p_i(\sqrt{\abs{g}}g^{ij}\p_j)\d x^- \nn\\
&&+\frac{r_0^2}{2\n+r_0^2 u} \d x^- \p_\ta(\sqrt{\abs{g}} \p_\ta) \d
x^-\Big\},
\end{eqnarray}
where
\begin{eqnarray}
u=1+\hat{\s}^2\ta_2^2\Xi_1^{-1}.
\end{eqnarray}
In the first order in $\ep$, $\p_\ta \d x^-=0$ as the endpoints in
$\ta$ are fixed. From the second term, which is proportional to $\d
r$ we get a constraint\footnote{The constraint is also obtained from
(\ref{eq:lm_j00}) and (\ref{eq:radii_j00}) with $\l:=1$ and
$\m:=\n$. }
\begin{eqnarray}
\n_\pm=\frac{2r_0^2}{9}\lt[-1-3\hat{\s}^2\ta_2^2\Xi_1^{-1}\pm\sqrt{1-3\hat{\s}^2\ta_2^2\Xi_1^{-1}}\rt].
\end{eqnarray}
$\n_+$ minimizes the action.

To find the spectrum we decompose the solution as
\begin{eqnarray}
\d x^I=\d \tilde{x}^I e^{-i\o\ta}Y_{l,\a}.
\end{eqnarray}
$Y_{l,\a}$ are four-dimensional spherical harmonics which satisfy
\begin{eqnarray}
\frac{1}{\sqrt{\abs{g}}}\p_i(\sqrt{\abs{g}}g^{ij}\p_j)Y_{l,\a}=-q_l
Y_{l,\a},~~q_l=l(l+2).
\end{eqnarray}
Due to the term $\sim \mcq^2\lt[(\p_{\ph_2}-\p_{\ph_3})\d x^I\rt]^2$
in the fifth line of the action (\ref{eq:act_pert}) the degeneracy
of the spherical harmonics is lifted. The spherical harmonics are
diagonalized as
\begin{eqnarray}
\lt(\frac{\p}{\p\ph_2}-\frac{\p}{\p\ph_3}\rt)^2Y_{l,\a}=-\a^2Y_{l,\a}.
\end{eqnarray}
The spectrum in the $X^a$, $(a=5,\cdots,8)$, directions is
\begin{eqnarray}
\o_a^2=1+\Xi_1^{-1}\mcq^2\a^2+\frac{1}{9}\lt(2+\sqrt{1-3\Xi_1^{-1}\hat{\s}^2\ta_2^2}\rt)^2q_l.
\end{eqnarray}
The radial direction and the null direction $X^-$ are coupled. The
equations of motion are
\begin{eqnarray}
&&s:=1-3\Xi_1^{-1}\hat{\s}^2\ta_2^2, \nn\\
&&\lt[\frac{8}{3}(\sqrt{s}-s)+\frac{1}{9}(2+\sqrt{s})^2q_l+\Xi_1^{-1}\mcq^2\a^2-\o^2\rt]\d
\tilde{r}-i\frac{\o}{r_0}\lt(\frac{6\sqrt{s}}{2+\sqrt{s}}\rt)\d
\tilde{x}^-=0, \nn\\
&&i\frac{\o}{r_0}\lt(\frac{6\sqrt{s}}{2+\sqrt{s}}\rt)\d
\tilde{r}+\frac{1}{r_o^2}\lt[q_l-\frac{9}{(2+\sqrt{s})^2}\o^2\rt]\d
\tilde{x}^-=0.
\end{eqnarray}
The spectrum is
\begin{eqnarray}
&&t:=\frac{4\sqrt{s}(2+\sqrt{s})}{3}+\Xi_1^{-1}\mcq^2\a^2, \nn\\
&&\o^2_\pm=\frac{t}{2}+\lt(\frac{2+\sqrt{s}}{3}\rt)^2q_l \pm
2\sqrt{\frac{t^2}{16}+s\lt(\frac{2+\sqrt{s}}{3}\rt)^2q_l}.
\end{eqnarray}
$\o^2_+$'s are positive definite while $\o^2_-$'s are positive
semidefinite. A zero mode occurs when $l=0$. The spectrum is
independent of the size $r_0$. The spectrum depends on the marginal
deformation parameters $\g$ and $\s$ as well as the axion-dilaton
parameters $\ta_1$ and $\ta_2$. There is no complex frequency. When
$\hat{\s}\neq0$, giant gravitons are not energetically favorable but
the spectrum of small fluctuations shows that the giant gravitons
are perturbatively stable. The result is consistent with the results
of \cite{Hamilton:2006ri,Avramis:2007wb}.
%
%
\section{Discussion}\label{sec:diss}
%
We have studied the Penrose limits of the complex marginal
deformation of $AdS_5\times S^5$ which incorporates the
$SL(2,\mathbb{R})$ symmetry of type IIB theory and have presented
the pp-wave geometries along the $(J,0,0)$ geodesic and along the
$(J,J,J)$ geodesic. We have shown that giant gravitons on the
$(J,0,0)$ pp-wave depend the parameter $\s$ as well as the
axion-dilaton parameters. Giant gravitons exist up to a critical
value of $\s$, which depends on the axion-dilaton parameters. The
spectrum of small fluctuations about the giant graviton solution is
obtained. The giant gravitons are energetically unfavorable but
perturbatively stable.

\vspace{1cm}


\end{document}